\documentclass[10pt,twocolumn,letterpaper]{article}

\usepackage{iccv}
\usepackage{times}
\usepackage{epsfig}
\usepackage{graphicx}
\usepackage{amsmath}
\usepackage{amssymb}
\usepackage[marginal]{footmisc}

\usepackage[breaklinks=true,bookmarks=false]{hyperref}

\iccvfinalcopy 

\def\iccvPaperID{****} 

\begin{document}

\title{SRDGAN: learning the noise prior for Super Resolution with Dual Generative Adversarial Networks}

 \author{
{Jingwei GUAN*, Cheng PAN*, Songnan LI and Dahai YU
}\\
TCL Corporation Research (Hong Kong)
}

\maketitle
\footnote{\noindent * They contributed equally to this work.}
\begin{abstract}
Single Image Super Resolution (SISR) is the task of producing a high resolution (HR) image from a given low-resolution (LR) image. It is a well researched problem with extensive commercial applications such as digital camera, video compression, medical imaging and so on.
Most super resolution works focus on the features learning architecture, which can recover the texture details as close as possible \cite{SRCNN, FSRCNN, ESRGAN,SRGAN}. 
However, these works suffer from the following challenges: 
(1) The low-resolution (LR) training images are artificially synthesized using HR images with bicubic downsampling, which have much richer-information than real demosaic-upscaled mobile images.
The mismatch between training and inference mobile data heavily blocks the improvement of practical super resolution algorithms.
(2) These methods cannot effectively handle the blind distortions during super resolution in practical applications. 
In this work, an end-to-end novel framework, including high-to-low network and low-to-high network, is proposed to solve the above problems with dual Generative Adversarial Networks (GAN).
First, the above mismatch problems are well explored with the high-to-low network, where clear high-resolution image and the corresponding realistic low-resolution image pairs can be generated.
Moreover, a large-scale \textbf{General Mobile Super Resolution Dataset, GMSR,} is proposed, which can be utilized for training or as a fair comparison benchmark for super resolution methods.
Second, an effective low-to-high network (super resolution network) is proposed in the framework.
Benefiting from the GMSR dataset and novel training strategies, the super resolution model can effectively handle detail recovery and denoising at the same time.
\end{abstract}

\section{Introduction}

\begin{figure}[t]
\begin{center}
\includegraphics[width=1.1\linewidth]{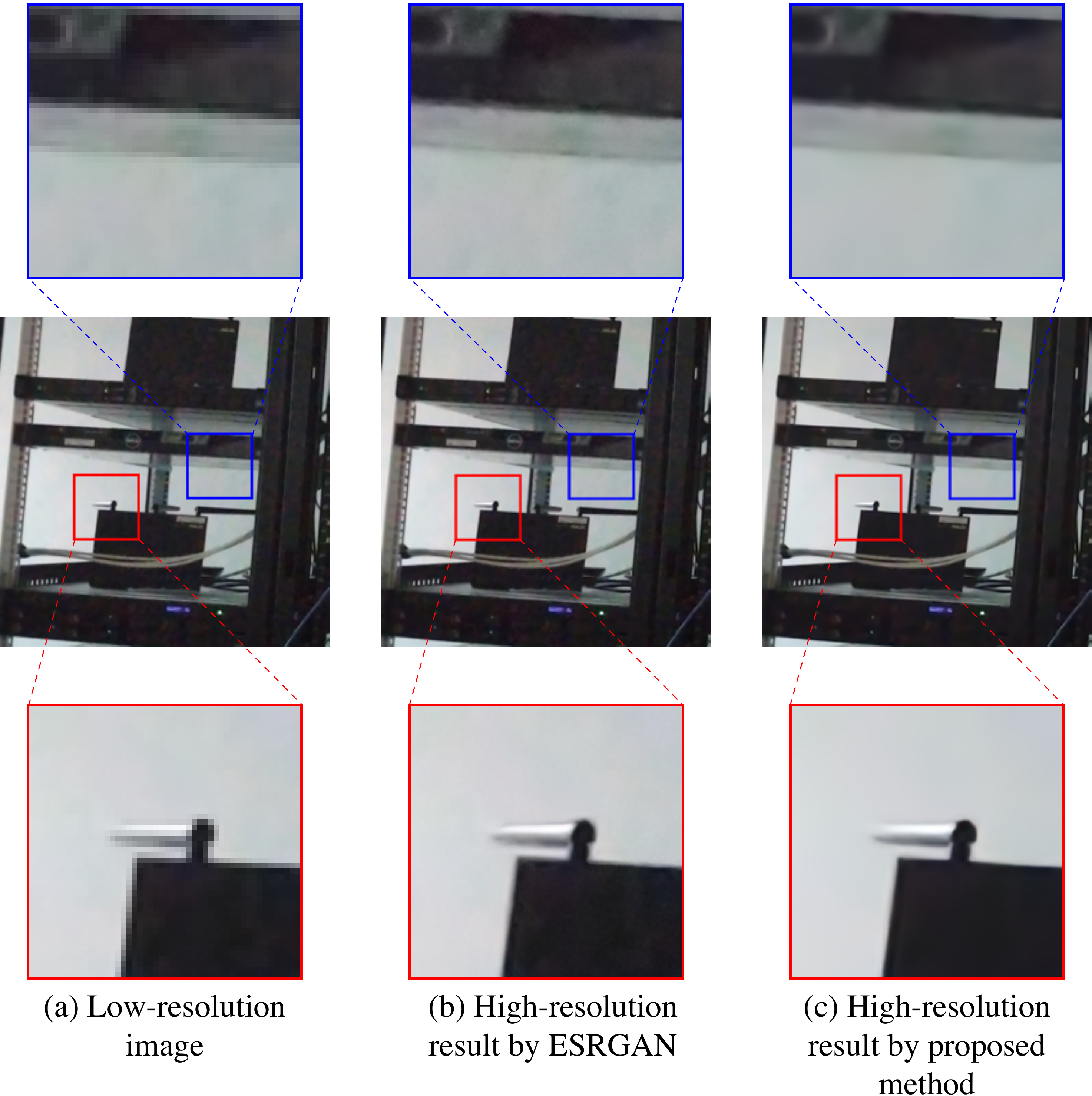}
\end{center}
\label{fig:motivation}
\caption{{\small
(a) Example of an mobile low-resolution image, which is utilized as the input of super resolution (SR) methods.
(b) Examples of super-resolution result via state-of-the-art method ESRGAN \cite{ESRGAN}. 
(c) Examples of SR result via the proposed method.
(Best viewed in color.)
}}
\end{figure}


Among all photographic equipment, mobile has developed as the most popular one because of its portability and evolving quality. 
Most cameras on mobile phones support 1x-10x zoom level change.
However, using optical zoom to achieve the large zoom-level change is not realistic.
Thus, digital zoom method, \emph{i.e.} super resolution, is extensively utilized in smart phones.

This paper targets on single image super resolution (SISR), which produces the high-resolution image based on a single low-resolution image. Various works have been done to recover the details and generate realistic high-resolution (HR) images.
Recently, learning based methods have shown great potential in solving low-level problems \cite{LLV1,LLV2,LLV3,LLV4}.
Especially, development of deep learning provides various learning-based solutions for super resolution \cite{SRCNN,FSRCNN, SRGAN, ESRGAN}.
Details and patterns are expected to be learned from large-scale data sets.
Among the learning-based methods, generative adversarial network (GAN)\cite{gan} achieved great progress. 
It has been proved that GAN-based methods \cite{SRGAN,ESRGAN} can generate plausible-looking natural details which are consistent with the human visual system (HVS).

However, mismatches exist between the data sets utilized by most existing learning-based methods and the real images captured by a mobile phone.
First, in most super resolution methods, low-resolution(LR) images in popular data sets, such as Set5, Set14, are simply downsampled by bicubic interpolation of the high-resolution image.
Each pixel of the LR image is obtained by weighted averaging the corresponding patch of the HR image. 
Thus the informative level of the training low-resolution image is much richer than $1/N^2$ of the high-resolution image, where N represents the downscale factor.
To sum, the training LR image is rich-informative.
While in the real case, each pixel of the mobile captured image is actually upscaled by demosaicing.
Thus, the information of mobile images is under-informative than its actual image size.
Therefore, the first mismatch between the training LR data and the mobile captured data is the informative level.
Second, the LR image in training data is clear and with no noise and distortions. 
While the LR image captured by mobile suffers from many distortions, such as noise and blur, as shown in Fig. \ref{fig:motivation} (a).

These mismatches between the downsampled LR training data and real LR data (mobile-captured data) heavily restrict the performance of the algorithm and lead to unsatisfactory results. 
As shown in Fig. \ref{fig:motivation} (b), the HR image generated by the state-of-the-art method ESRGAN \cite{ESRGAN} with DIV2K as training data cannot handle real mobile images well.





Based on the observation, an end-to-end framework is proposed which consists of two parts.
First, a high-to-low network $\Phi_{H2L}$ is trained to generate realistic HR/LR image pair for training super resolution models. 
Benefiting from the $\Phi_{H2L}$, a new dataset, \textbf{GMSR}, is generated which provides training and testing sets accompanied with references of different similarity levels in terms of noise reduction, texture, color, illumination, view point, etc.
This dataset can promote fair comparison and support further research on the SR problems in general.
Second, a low-to-high network $\Phi_{L2H}$ is trained to produce the super resolution image.
Novel training methods were utilized to reproduce better results.
As shown in Fig. \ref{fig:motivation} (c), the HR image generated by our method can handle the noise better than ESRGAN \cite{ESRGAN} as illustrated by the blue patch.
Moreover, better details can be reproduced as shown in the red patch.
To sum up, the proposed super resolution method can effectively deal with the real distortions and restore fine details.

The contributions of our work can be summarized into three categories.
First, a general SR problem is explored which straightly face the mismatch between real mobile images and commonly utilized training data.
A H2L network is proposed to solve the problems effectively.
Second, an end-to-end framework is proposed, which includes two parts, high-to-low network (H2L) and super-resolution network (L2H). 
During training, dual Generative Adversarial Network (GAN) is utilized to optimize the parameters.
Besides, novel training strategies, such as using nearest neighbour downsampling method to generate training pairs, were explored to further improve the performance.
We demonstrate the visual improvement, fine reproduced details and denoising effect during super resolution of the proposed dual-generative network by extensive empirical studies.
Third, a large-scale General Mobile Super Resolution dataset, \textbf{GMSR}, is generated by high-to-low network.
This dataset can support further research on the super resolution domain, and utilized either for training or as a fair benchmark.

\section{Related Work}

Single image super resolution (SISR) is an important topic and has been developed for a long time. 
Early methods \cite{intp1,intp2,intp3,intp4} that based on the interpolation theory can be very fast, however usually yield over-smooth results. 
Methods rely on neighbor embedding and sparse representation \cite{ne, ne2, anch, anch2, sc, ksvd} targeted on learning the mapping between LR and HR. 
Some example-based approaches used image self-similarity property to reduce the amount of training data needed \cite{self1,self2,self3}, and increased the size of the limited internal dictionary \cite{self4}.

With the development of deep learning technology, methods based on convolution neural network have shown great potential in solving super resolution problems. 
Dong \emph{et al.} \cite{SRCNN} first proposed an end-to-end convolutional neural network to learn the mapping between HR and LR. 
Kim \emph{et al.} \cite{kim1} improved the reconstruction accuracy by using very deep convolutional network and residual learning, and they further utilized recursive structure and skip-connection to improve performance without introducing new parameters \cite{kim2}. 
To increase inference efficiency, methods with low-resolution images as input \cite{laplacian,FSRCNN,sub_pixel,edsr} were proposed.
In \cite{laplacian}, the authors proposed the Laplacian Pyramid network structure to reduce the computational complexity and improve the performance. 
Dong \emph{et al.} \cite{FSRCNN} and \cite{sub_pixel} introduced different deconvolution methods to upscale the LR to HR. 


Most recently, GAN based super resolution methods have been proposed and achieved great progress.
Ledig \emph{et al.} \cite{SRGAN} first applied GAN to super resolution task and got highly photo-realistic results. 
ESRGAN \cite{ESRGAN} improved the performance by modifying the network structure and loss based on SRGAN. 
Network conditioning was used in \cite{sftgan} to combine the category prior with the generative network to get more realistic textures. 
Yuan \emph{et al.} \cite{cycle_gan} developed unsupervised learning with GAN, where unpaired HR-LR data were utilized.
Until now, most super resolution methods use the bicubic downsampled LR image as training data. 
However, since bicubic downsampling is quiet different from the degradation in real-world, it makes the trained model not effective to the real-world cases. 
Inspired by ESRGAN \cite{ESRGAN}, We followed its main structure and proposed a deeper network named the low-to-high(L2H) network to solve the more complicated real-world cases, and build a large-scale General Mobile Super Resolution Dataset \textbf{GMSR}, which is potentially applicable to other image reconstruction methods. 
Bulat \emph{et al.} \cite{degradationface} adopted a degradation network to generate LR image first, then use it to train the model and get good result. 
However, they focus on face category, which has much less complexity and diversity than general image. 
Hence, we followed their idea and extended their work to the general image super resolution. 
We proposed a high-to-low(H2L) network to learn the degradation from HR image to LR image, so that the 'realistic' LR image can be generated once we have the HR image. 
Zhao \emph{et al.} \cite{degradation2} proposed a degradation network to generate 'realistic' LR image and use it to train the degradation and SR construction network. However, they only use one discriminative network to train the degradation network. 
Different from them, two discriminative networks were utilized to train H2L and L2H network respectively in this work.
Moreover, the two networks were jointly optimized as well.

\begin{figure*}[t]
\begin{center}
\includegraphics[width=1\linewidth]{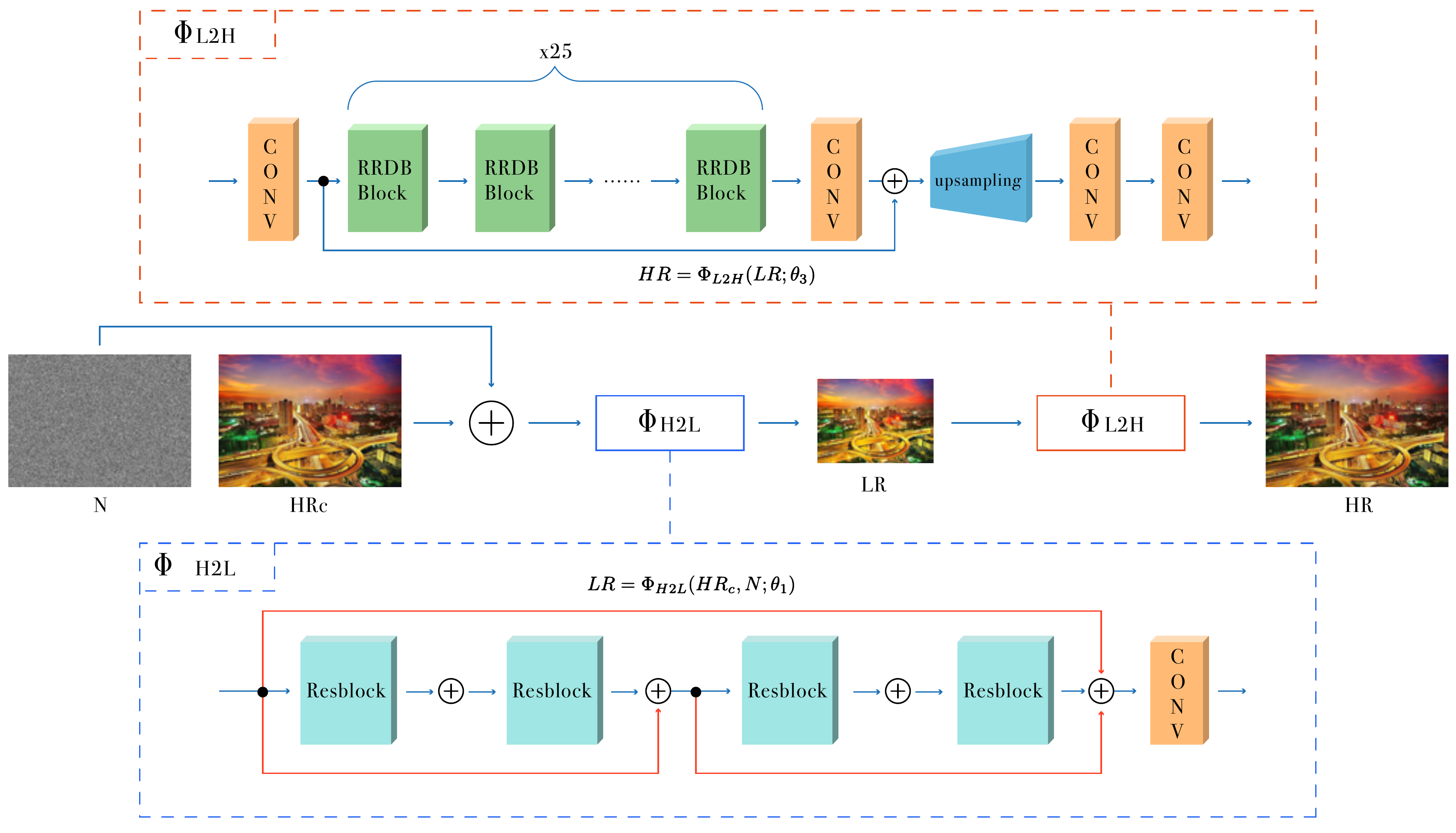}
\end{center}
\caption{Pipeline of the proposed SRDGAN framework, including $\Phi_{H2L}$ and $\Phi_{L2H}$. 
The high-to-low network $\Phi_{H2L}$ is proposed to generate realistic HR/LR image pairs.
$N$ is set as the input of $\Phi_{H2L}$, which can add the randomness when generating realistic distorted LR image.
The low-to-high network $\Phi_{L2H}$ is proposed to produce the super resolution image.
The detailed architecture of $\Phi_{H2L}$
and $\Phi_{L2H}$ are illustrated in the top and bottom row respectively. (Best viewed in color.)
}
\label{fig:pipeline}
\end{figure*}

\begin{table*}
\begin{center}
\caption{~~Notations used in the paper. }
\begin{tabular}{c|c}
\hline
$\Phi_{H2L}$ & High-to-low Network \\
$I_{pair}$=\{HR$_c$, HR$_n$\} & Pixel-level corresponding image pairs, including clear image HR$_c$ and noisy image HR$_n$ \\
LR$_n$ & Low-resolution image downsampled from HR$_n$\\
$I_n$ & Captured large-scale noisy images by phone. No corresponding clear images\\ $I_{nc}$ & Cropped LR image from $I_n$\\
\textbf{GMSR} & Proposed general mobile super resolution dataset\\
$I_{sim}$ & 
Image pairs in dataset \textbf{GMSR}\\
HR$_c$ & Clear high-resolution images as input of $\Phi_{H2L}$ \\
N &  Random noise added to HR$_c$\\
$\Phi_{L2H}$ & Low-to-high Network, i.e. super-resolution network \\
\hline
\end{tabular}
\end{center}
\label{table:notation}
\end{table*}

\section{Proposed Framework}
The proposed framework aims at solving real-world super resolution problems.
In the framework, a high-to-low network is first proposed to generate realistic image pairs, which is trained with GAN using real-word images captured by mobiles. (Section \ref{sec:H2L}).
Secondly, a super resolution method with a novel structure and training strategy is explored (Section \ref{sec:L2H}).
Besides, new datasets are captured or generated, which are also introduced in this section.

\subsection{Datasets}
Three novel datasets were captured and generated to train the proposed framework, including $I_{pari}$, $I_n$ and \textbf{GMSR}.
$I_{pair}$ and $I_n$ were captured to train the $\Phi_{H2L}$ network.
Dataset \textbf{GMSR} is simulated by $\Phi_{H2L}$ and utilized to train the super resolution model $\Phi_{L2H}$.
Details of the network as given as follows.

\begin{enumerate}

\item 
In order to learn the noise distribution of real world images, we applied smart phone to capture the dataset, named $I_{pair} = \{HR_c, HR_n\}$. 
Blackberry Key2 \cite{key2} is selected to do this task duo to its optical module is very representative for many general common mobile devices .  
The resolution of images in $I_{pair}$ are $4032\times 3024$.
It contains $447$ pairs of images, which are randomly divided into training, validation and test set with number 400, 27 and 20 respectively.
Each pair contains a noisy image $HR_n$ and a corresponding clear image $HR_c$.
To generate the clear image $HR_c$, a burst of 20 noisy images were captured with the mobile phone fixed with a tripod.

A multi-frame denoising method was used to generate a clear high-resolution image $HR_c$ by adaptively fusing the 20 noisy images.
Since any one of the 20 noisy images can be used for generating training image pairs.
At most 20 pairs of images can be generated for each scene.
During training, $LR_n$ is down-sampled from $HR_n$, and constitutes an image pair with $HR_c$.

\item 
Dataset $I_n$ contain 206 captured noisy image by Blackberry key2 \cite{key2} with resolution $4032\times3024$. Contents of the images are different from the ones in $I_{pair}$. We randomly cropped $I_n$ to the size of $LR_n$, and input it to the D network as reference to help the high-to-low network to learn the actual noise distribution.
\item 
General Mobile Super resolution \textbf{GMSR} is generated using $\Phi_{H2L}$ and contains 1153 image pairs.
$I_{sim}$ represent the image pairs.
A large amount of clear images are collected from the Internet and regarded as HR images.
The corresponding low-resolution image is generated by $\Phi_{H2L}$, and the details are introduced in Sec. \ref{sec:H2L}.
The simulated dataset $I_{sim}$ is used to train $\Phi_{L2H}$.


\end{enumerate}

\begin{figure*}[t]
\begin{center}
\includegraphics[width=1.0\linewidth]{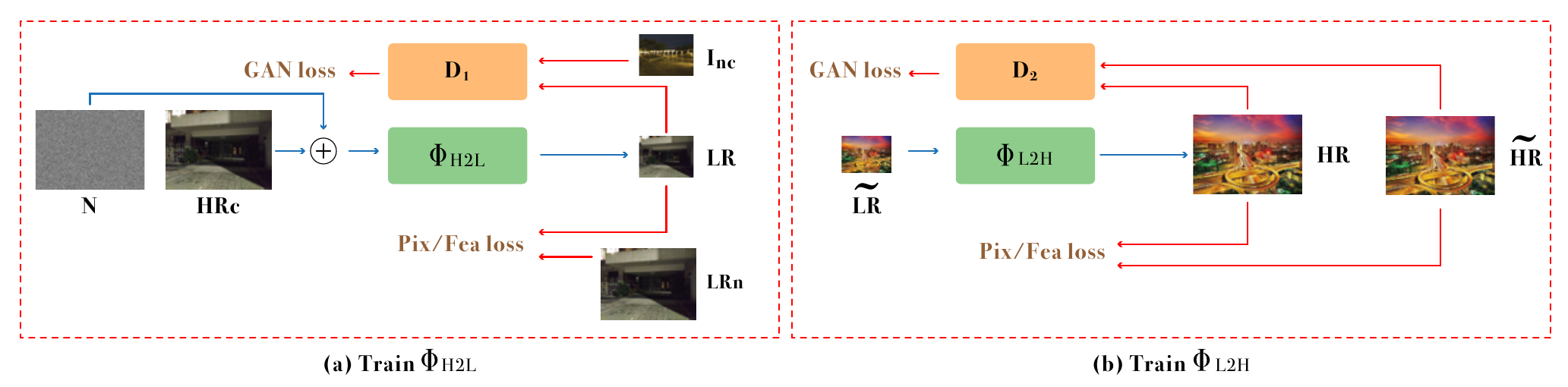}
\end{center}
\caption{Training procedure of the proposed framework with dual GAN.
During training, three kind of losses are used, including pixel/feature loss(Pix/Fea loss) and GAN loss.
(a) introduces the training process of $\Phi_{H2L}$, where dataset
$I_{pair}=\{HR_c, HR_n\}$ and $I_n$ are utilized.  
$LR_n$ is the downsampling result of $HR_n$ by nearest neighbour.
$I_{nc}$ is cropped from $I_n$ for the GAN loss.
(b) is the training procedure of $\Phi_{L2H}$. $\tilde{LR}$ and $\tilde{HR}$ represent the image pairs utilized for training, which are obtained from dataset \textbf{GMSR} and DIV2K\cite{div2k}. (Best viewed in color)
}
\label{fig:TrainWhole}
\end{figure*}

\subsection{High-to-low Network} \label{sec:H2L}
The biggest problem faced by SR is lacking realistic datasets, \emph{i.e.} noisy low-resolution image and the corresponding high-resolution image. 
The difficulty of generating such image pairs is mainly in two-fold.
One is to model and simulate distortions in LR images, 
the other is to generate image pairs with pixel-wise correspondence.


To solve the problem, a high-to-low network $\Phi_{H2L}$ is proposed.
Notations used in this paper are summarized in Table \ref{table:notation}.
LR image is generated based on a clear $HR_c$ as follows:

 \begin{equation}\label{equ:H2L}
 \begin{aligned}
 LR = \Phi_{H2L}(HR_c, N ; \theta_1)
 \end{aligned}
 \end{equation}
where $HR_c$ is a clear high-resolution image, 
the generated low-resolution image $LR$ is supposed to have similar noise distribution with real-world image.
Gaussian random noise $N$ is added to simulate the randomness of the distortions in LR. The mean and standard deviation of the Gaussian noise is 0 and 0.05, respectively. The two values can be adjusted according to the distortion level of different mobiles.
The noise transition between Gaussian distribution and that in real-world images are accomplished by $\Phi_{H2L}$ and learned through GAN.






\subsubsection{Model Structure}
The structure of the $\Phi_{H2L}$ network is shown in Fig. \ref{fig:pipeline}, which consists of four Resblocks and one convolutional layer.
Each of the Resblock consists of two convolution layers and two activation functions.
In order to further build the global connection \cite{ECCV_2018}, multiple shortcut connections is added as indicated by the red lines.
The final convolution layer aims at decreasing the resolution by setting the downscale factor as the stride. 
$\Phi_{H2L}$ is able to model the image distortion of real mobile images caused in the image capturing and processing pipeline.



\subsubsection{Training Procedure of $\Phi_{H2L}$}
\label{sec:trainH2L}
The training strategy of $\Phi_{H2L}$ is summarized in Fig. \ref{fig:TrainWhole}.
Inputs of $\Phi_{H2L}$ are $HR_c$ and random noise $N$, and the output is $LR$. 
During the training process, dataset $I_{pair}=\{HR_n, HR_c\}$ and $I_n$ are used.
In order to get the ground-truth of $LR$, noisy image $HR_n$ is down-sampled to generate $LR_n$, which is used to calculate the pixel and feature losses.
However, the noise distribution of $LR_n$ is not consistent with $HR_n$ after the down-sampling.
Therefore, Generative Adversarial Network (GAN) is used to further learn the noise distribution of mobile images.
The loss function and down-sampling strategy are introduced as follows.

\textbf{Loss function}:
Pixel loss, feature loss (i.e. perceptual loss) and GAN loss are used to train $\Phi_{H2L}$.
 \begin{equation}\label{equ:lossH2L}
 \begin{aligned}
loss_{H2L} = \alpha_1(loss_{H2L}^p)+ \alpha_2(loss_{H2L}^f)+\alpha_3(loss_{H2L}^{GAN})
 \end{aligned}
 \end{equation}
where $\alpha_1$, $\alpha_2$ and $\alpha_3$ are weighting parameters.

Pixel loss $loss_{H2L}^{p}$ aims at pixel-wisely comparing the image difference between $LR_n$ and the network output $LR$, as 
 \begin{equation}\label{equ:H2L_pixelLoss}
 \begin{aligned}
loss_{H2L}^{p}= \frac{1}{n_{LR}^p} \sum_{k= 1}^{n_{LR}^p}(|| LR(k)-LR_n(k)||_2^2)
 \end{aligned}
 \end{equation}
where $n_{LR}^p$ corresponds to the number of pixels in $LR$.

Feature loss aims at comparing the difference between high-level features of two images, which has been proved to be effective in many previous works\cite{SRGAN,ESRGAN,He_2016_CVPR,Perceptual_loss}. Inspired by these works, VGG\cite{vgg} is utilized as feature extraction network $\varphi$:  
 \begin{equation}\label{equ:H2L_featureLoss}
 \begin{aligned}
loss_{H2L}^{f}= \frac{1}{n_{LR}^f} \sum_{k= 1}^{n_{LR}^f}(|| \varphi (LR)(k)-\varphi(LR_n)(k) ||_1)
 \end{aligned}
 \end{equation}
$\varphi (LR)$ is the extracted high-level features of input image LR by $\varphi$. $k$ represents the index in the extracted features.


However, $LR_n$ is generated by downsampling mobile images.
The noise distribution in $LR_n$ cannot be well maintained.
Thus GAN is used to learn the distribution by discriminating unpaired images, $I_{nc}$ and $LR_n$, where $I_{nc}$ is the low-resolution image randomly cropped from $I_n$.

 \begin{equation}\label{equ:H2L_GANLoss}
 \begin{aligned}
 loss_{H2L}^{GAN} = \frac{1}{n_{LR}^{GAN}} \sum_{n=1}^{n_{LR}^{GAN}}{-logD_{\theta_2}(\Phi_{H2L}(HR_n,N))}\\
 \end{aligned}
 \end{equation}
where $n_{LR}^{GAN}$ is the number of batch size, $D_{\theta_2}$ is the discriminative model with parameters $\theta_2$.
The goal of $D_{\theta_2}$ is to well distinguish the generated LR image and real image $I_{nc}$.
$D_{\theta_2}$ is trained iteratively with the generative model $\Phi_{H2L}$ as shown in Equ. \ref{equ:GAN}.
Thus, better $D_{\theta_2}$ is trained to distinguish results generated by $\Phi_{H2L}$, while better $\Phi_{H2L}$ is trained to "confuse" the discriminator.

 \begin{equation}\label{equ:GAN}
 \begin{aligned}
min_{\theta_1}max_{\theta_2} &\mathbb{E}[log D_{\theta_2}(I_{nc})]+\\
        &\mathbb{E}[log(1- D_{\theta_2}(\Phi_{H2L}(HR_N,N)))]
 \end{aligned}
 \end{equation}

\textbf{Downsampling strategy:}
Bicubic interpolation is used as the down-sampling method to generate training LR image in most previous works \cite{ESRGAN,SRGAN,SRCNN,FSRCNN}.
However, images downsampled by bicubic interpolation contain much more information than real-world images captured by mobiles.

To be specific, the luminance of a pixel in the low-resolution image is a weighted average of the luminance of its surrounding pixels during the bicubic interpolation. 
Therefore, rich information is contained in the bicubic-downsampled images.
However, in real-world image cases, the multi-channel color images captured by mobile are actually interpolated by single-channel image using the demosaic of the image signal processing(ISP).
The information level of real-world images and bicubic downsampled images are unbalanced. Therefore, 
models trained with bicubic-downsampled image pairs certainly cannot work well on real-world images.

In this work, nearest neighbour is adopted as the down-sampling method to better simulate the real-world ill-informative situation of mobile images.

\subsection{Low-to-high Network} 
\label{sec:L2H}
Network $\Phi_{L2H}$ is proposed to implement super resolution based on the low-resolution image $LR$ with parameter $\theta_3$ as in Equation \ref{equ:L2H}.
 \begin{equation}\label{equ:L2H}
 \begin{aligned}
 HR = \Phi_{L2H}(LR ; \theta_3)
\end{aligned}
\end{equation}
\subsubsection{Network Structure}

The structure of $\Phi_{L2H}$ is shown in Fig. \ref{fig:pipeline}. Following ESRGAN \cite{ESRGAN}, RRDB block is used as the basic network unit. Each RRDB block contains 3 residual blocks(RDB), and each residual block contains 5 convolution layers without BN layers in a dense structure. The basic unit is not limited to RRDB block, other basic block like residual block and dense block can work as the basic unit in the proposed network structure. However, since the complexity of the real-world super resolution task is high, a deeper structure with 25 RRDB blocks is proposed to model the more complex task.


\subsubsection{Training Procedure of $\Phi_{L2H}$}
The training strategy of $\Phi_{H2L}$ is summarized in Fig. \ref{fig:TrainWhole}. Significant improvements have been achieved by new training strategies, i.e. using simulated realistic training data and nearest neighbour as the down-sampling method.
Benefiting from these, the SR model is capable of dealing with the real world image distortions.

The loss function  of $\Phi_{L2H}$ is similar to that of $\Phi_{H2L}$, which is described in Section \ref{sec:trainH2L}.
 \begin{equation}\label{equ:loss_L2H}
 \begin{aligned}
loss_{L2H} = \alpha_4(loss_{L2H}^p)+ \alpha_5(loss_{L2H}^f)+\alpha_6(loss_{L2H}^{GAN})
 \end{aligned}
 \end{equation}
where $\alpha_4$, $\alpha_5$ and $\alpha_6$ are weighting parameters.
The difference between $\Phi_{H2L}$ and $\Phi_{L2H}$ is mainly at training data. 
For $\Phi_{L2H}$, two categories of training samples are utilized. The most important one is $I_{sim}$ from dataset GMSR that simulates real-world distortions. Furthermore, the popular dataset DIV2K \cite{div2k} are also utilized with nearest neighbour as the downsampling method, to further increase the data diversity, and encourage the model to perform well both in real-world cases and traditional benchmarks.



Finally, global fine-tuning is performed on the whole network structure including $\Phi_{H2L}$, $\Phi_{L2H}$ and the discriminative networks, to jointly optimize the performance. 

\begin{figure*}[t]
\begin{center}
\includegraphics[width=1\linewidth]{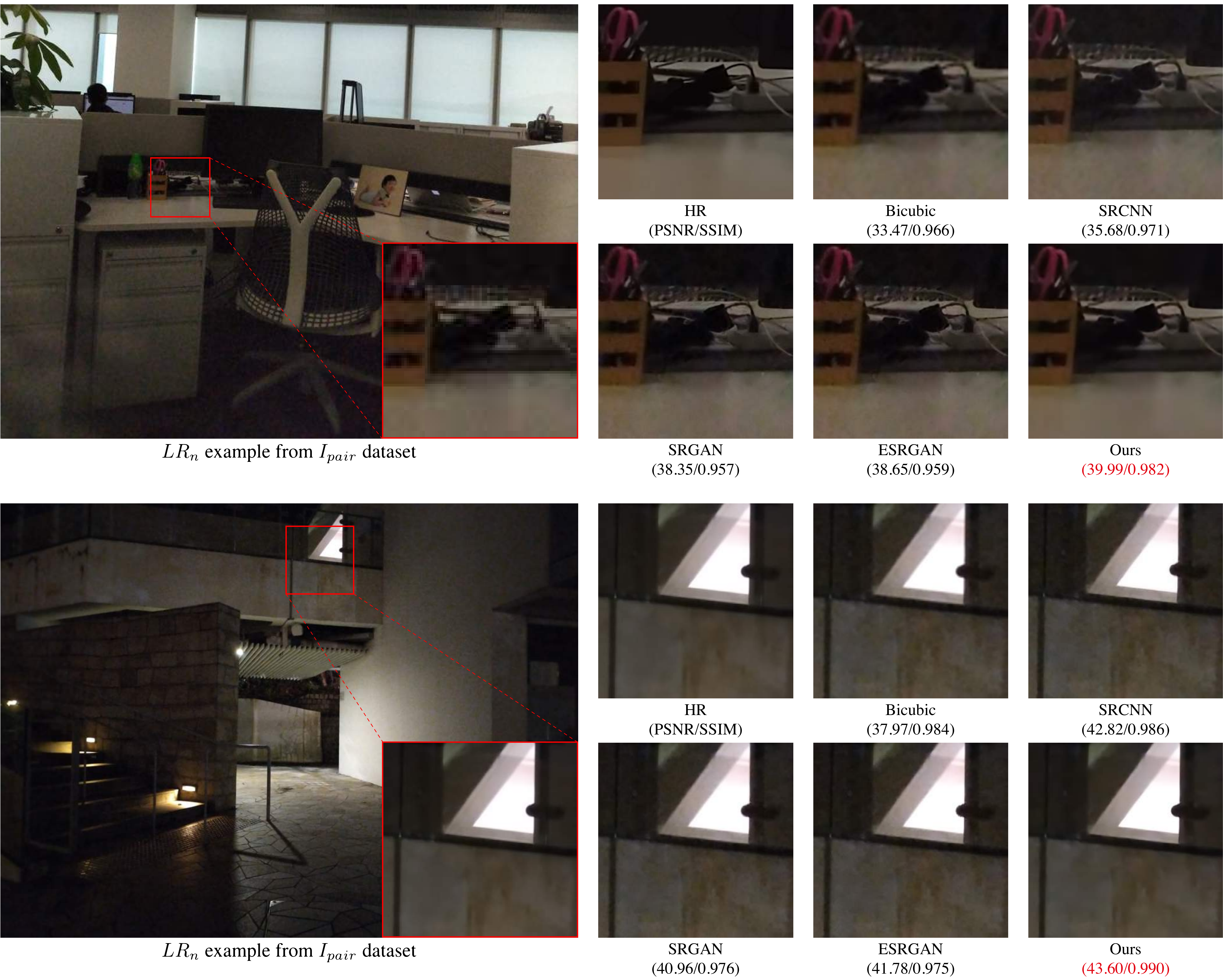}
\end{center}
   \caption{Subjective comparison between the proposed method some state-of-the-art methods. Left is the mobile low-resolution image example and a detail patch. Right is the super-resolution result of different methods and high-resolution ground-truth (HR). (Best viewed in color)
   }
\label{fig:Result_main}
\end{figure*}
\section{Experiments}
Experiments were conducted to evaluate the effectiveness of the proposed framework.
First of all, the SR results trained based on the proposed framework are presented. 
Second, effectiveness of each component is evaluated, including results generated by $\Phi_{H2L}$ and the novel training strategies and structure utilized in $\Phi_{L2H}$.

The parameters for training the proposed method are as follows.
Batch-size is 16 with training patch of size $192\times192$.
Following ESRGAN\cite{ESRGAN}, the learning rate is initialized with $1\times10^{-4}$, and halved at [50k, 100k, 200k, 300k] iterations.
Adam\cite{Adam} optimizer is utilized with $\beta_1 = 0.9$, $\beta_2 = 0.999$ without weight decay.

\subsection{Super Resolution Results}
\begin{figure*}[t]
\begin{center}
\includegraphics[width=0.95\linewidth]{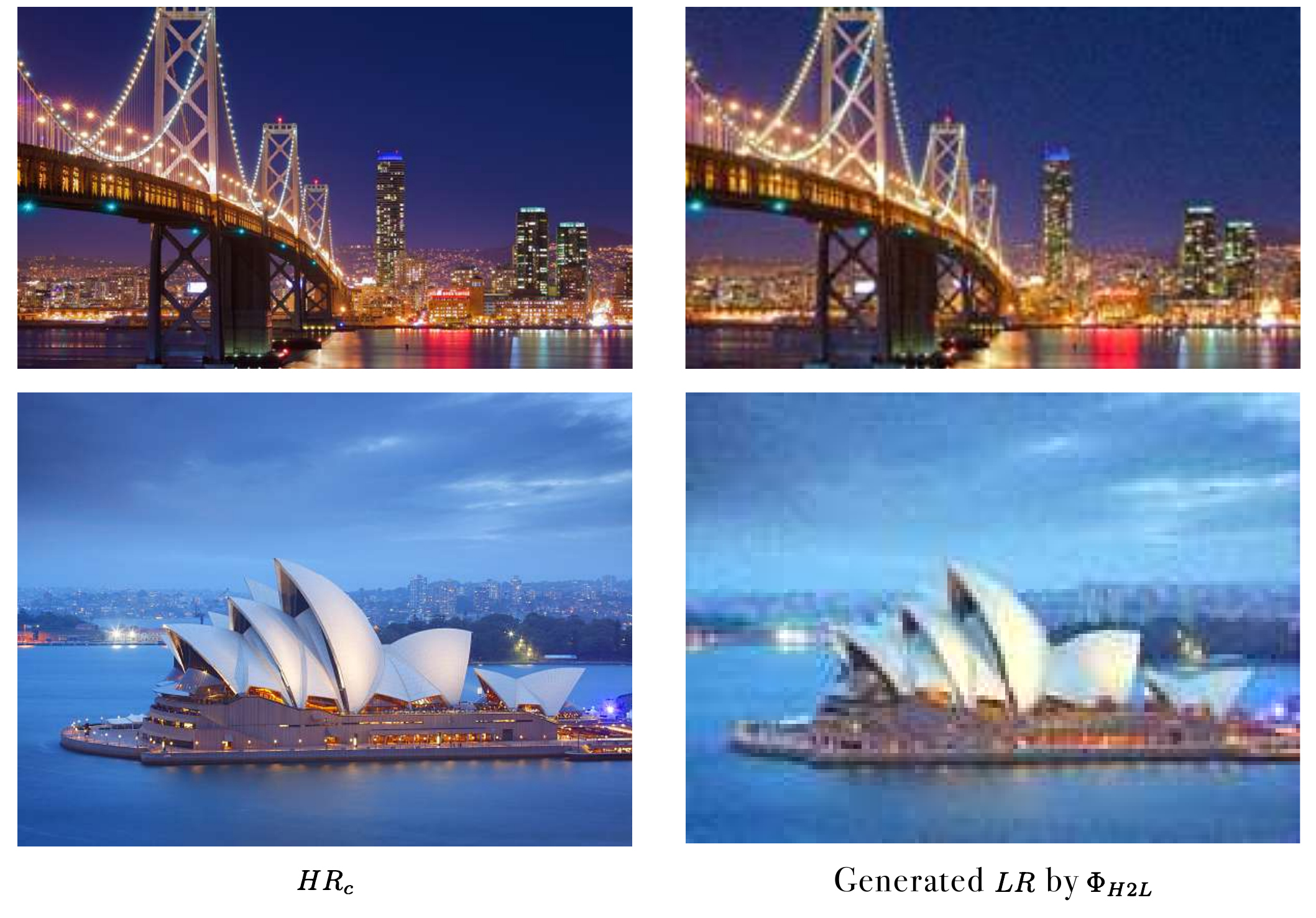}
\end{center}
   \caption{Demonstration of the image pairs generated by $\Phi_{H2L}$. Please noted that the resolution of $HR_c$ and Generated $LR$ are different. (Best viewed in color)}
\label{fig:ResultsOFH2L}
\end{figure*}

Subjective and quantitative experiments were conducted to evaluate the performance of the proposed method.
Table \ref{tabel:main} shows the quantitative evaluation results of the popular and state-of-the-art GAN-based super resolution methods with an upscale factor $4$. 
For SRCNN \cite{SRCNN} and SRGAN \cite{SRGAN}, the Set5 and Set14 results are copied from the original papers, and the results of $I_{pair}$ dataset are generated based on the model provided by their authors. 
Results of ESRGAN \cite{ESRGAN} are also calculated by the code provided by the author.
Since ESRGAN did not have quantitative result in the paper, we download their GAN-based model to test on Set5, Set14 and $I_{pair}$.
To evaluate the performance of the proposed method, both general datasets (set5\cite{set5}, set14\cite{set14}) and the mobile-specific dataset ($I_{pair}$) are utilized.
PSNR and SSIM \cite{SSIM} are adopted for measuring the effectiveness of the proposed method, where higher value indicates better performance.
The best performing results are shown in bold.

It is shown that the proposed method outperforms the other state-of-the-art methods in terms of SSIM\cite{SSIM} in all circumstances, demonstrating its effectiveness on reconstructing contrast, luminance and structure.
Especially, the improvements achieved on Set $I_{pair}$ is significant, more than 1 dB improvement in PSNR and 0.01 improvement in SSIM showing its good performance on mobile images.

Subjective comparison results were provided as well as shown in Fig. \ref{fig:Result_main}.
real-world mobile images are utilized.
PSNR and SSIM results are presented at the bottom for reference, which are evaluated on Y channel. 
In many situations, mobile images suffer from noise and detail missing.
Most methods cannot handle these cases well, including bicubic, SRCNN \cite{SRCNN}, SRGAN \cite{SRGAN}, ESRGAN \cite{ESRGAN}.
The noise is still very obvious, and even increased in SR results.
Instead, our method can deliver good results by removing all the noise and enhancing the details.

 \begin{table}[htb!]\caption{Quantitative evaluation of SR algorithms(PSNR/SSIM)}\label{tabel:main}
	\begin{center}
	\vspace{-0.2cm}
		\begin{tabular}{c|c c c}
			\hline\hline
			Method&Set5&Set14&Set $I_{pair}$(test)\\
			\hline
			Bicubic&28.42/0.8104&26.00/0.7027&35.85/0.9674\\
			SRCNN&30.07/0.8627&\textbf{27.50}/0.7513&39.09/0.9716\\
			SRGAN&29.40/0.8472&26.02/0.7397&38.43/0.9596\\
			ESRGAN&\textbf{30.55}/0.8677&26.39/0.7246&38.55/0.9572\\		
			Ours&30.37/\textbf{0.8766}&27.35/\textbf{0.7767}&\textbf{39.93/0.9782}\\
			\hline	\hline
		\end{tabular}
	\end{center}
\end{table}

\begin{figure}[h!t]
\begin{center}
\includegraphics[width=1.0\linewidth]{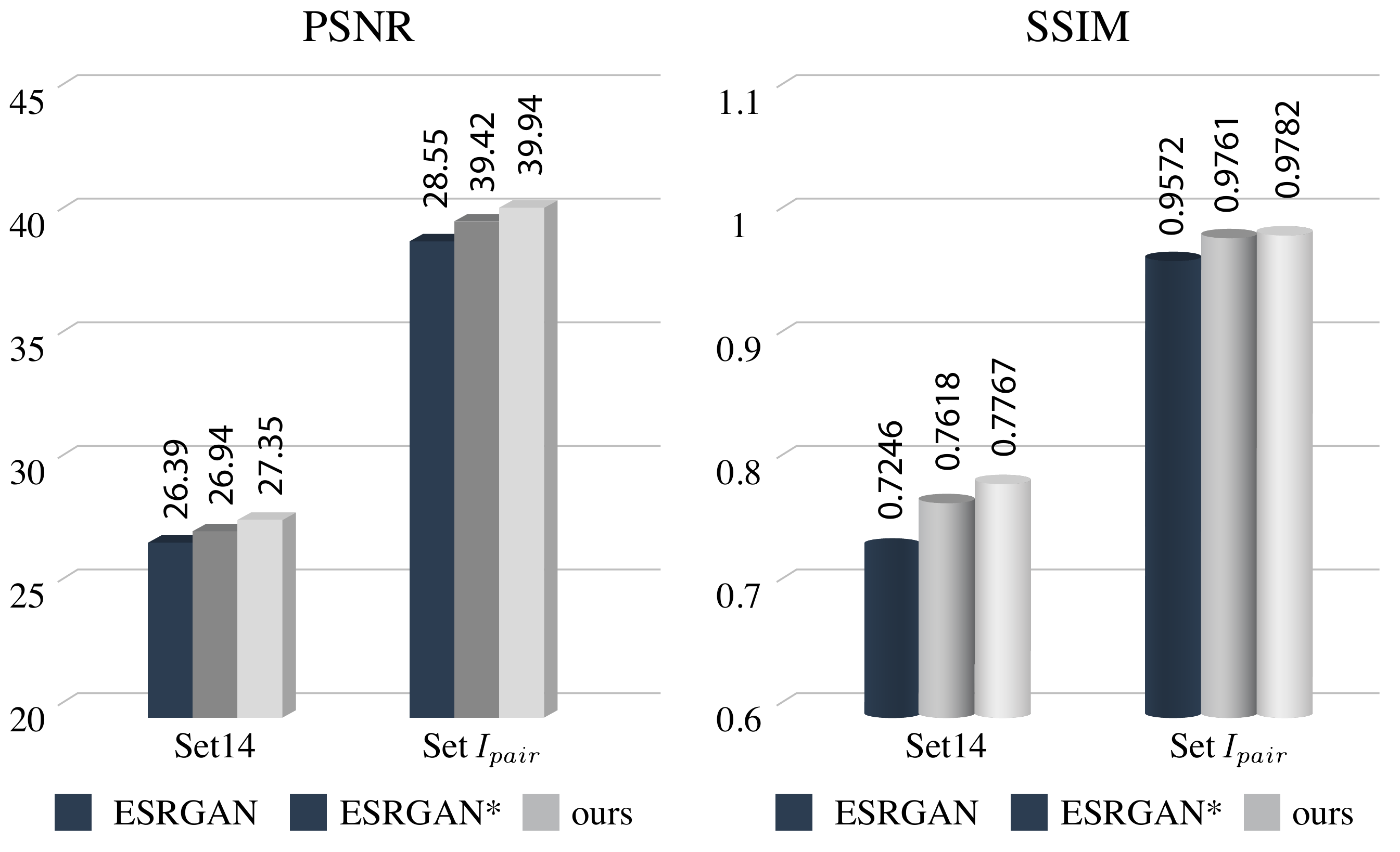}
\end{center}
   \caption{Comparison result of ESRGAN and the proposed method}
\label{fig:Result_structure}
\end{figure}

\subsection{Results of $\Phi_{H2L}$}
With $\Phi_{H2L}$, we aim to simulate LR images that is a better ensemble of real mobile images and to generate a large-scale dataset.
The image pairs generated by $\Phi_{H2L}$ are shown in Fig. \ref{fig:ResultsOFH2L}, including high-resolution input and corresponding low-resolution output.
It can be seen that the noise distribution of the generated LR images are consistent with mobile images.
The simulated noise levels of different position and on different objects are similar with the real ones.
Moreover, the color is well remained in the generated low-resolution images.
\vspace{-0.2cm}
\subsection{Exploring Training Strategies for $\Phi_{L2H}$}

Experiments were conducted to evaluate the effectiveness of the new dataset, new training strategies and the proposed model structure as shown in Fig. \ref{fig:Result_structure}.
'ESRGAN' represents the original ESRGAN method with GAN-based structure and DIV2K\cite{div2k} (bicubic downsampling) as training data.
'ESRGAN*' represents the new-trained method with the same model structure but new datasets for training, including DIV2K\cite{div2k} (Nearest neighbour downsampling) and GMSR.
'Ours' is the proposed method with the new model structure and new datasets for training.
Comparison between 'ESRGAN' and 'ESRGAN*' can evaluate the effectiveness of new training datasets.
While performance comparison between 'ESRGAN*' and 'Ours' evaluates the impacts in the model structure.

PSNR and SSIM \cite{SSIM} are utilized, where larger value represents better performance.
The test was conducted on both general dataset set14\cite{set14} and the mobile image dataset Set $I_{pair}$.
It can be seen that the 'ESRGAN*' outperforms 'ESRGAN' by changing training data. 
It proves that the new dataset(\textbf{GMSR}) and new training strategy (nearest neighbour) can be easily generalized to other model structures and improve the performance on both general data and mobile data.
Moreover, 'Ours' further improve the performance compared with 'ESRGAN*'.
It proves the effectiveness of the model structure in $\Phi_{L2H}$.

\section{Conclusion}
In this work, a novel framework SRDGAN is proposed to solve the noise prior super resolution problem with dual generative adversarial network. A general degradation network H2L is proposed and is able to learn the noise prior of real-world LR image. By using our proposed training strategy, the H2L network is trained to generate the 'realistic' LR image paired with the HR image, and A large-scale benchmark general mobile super resolution dataset, \textbf{GMSR}, is generated from it. The \textbf{GMSR} can help super resolution methods to be well applied in the real-world images. Meanwhile, a super resolution network L2H is proposed, which contains new structure and training strategies(i.e.using simulated realistic as training data and using the nearest neighbour as the down-sampling method). Especially, these training strategies are proved to be effective for other SR model, such as ESRGAN, and improve the performance.

{\small
\bibliographystyle{ieee}
\bibliography{egbib}
}

\end{document}